\documentclass[journal]{IEEEtran}
\usepackage{amssymb}
\usepackage{amsmath}
\usepackage{bm}
\usepackage{graphicx}
\usepackage{epstopdf}
\usepackage{algpseudocode}
\usepackage{algorithm}
\usepackage{array}
\usepackage{multirow}
\usepackage{cite}
\usepackage{amsthm}

\begin{document}
\title{User-Centric Multiobjective Approach to Privacy Preservation and Energy Cost Minimization in Smart Home}

\author{Hsuan-Hao Chang, Wei-Yu~Chiu,~\IEEEmembership{Member,~IEEE}, Hongjian Sun,~\IEEEmembership{Senior Member,~IEEE}, and Chia-Ming Chen

\thanks{This work was supported in part by the Ministry of Science and Technology of Taiwan under Grant MOST 107-2221-E-007-105, in part by
 the U.K. EPSRC under Grant EP/P005950/1 (TOPMOST Project), and in part by the European Union’s Horizon 2020 Research and Innovation Staff Exchange (RISE) Programme Marie Sk{\l}odowska-Curie Actions (MSCA) under Grant 734325 (TESTBED Project). (Corresponding author: Wei-Yu Chiu.)}
\thanks{H.-H. Chang is with the Department of Electrical Engineering, Yuan Ze University,
Taoyuan 32003, Taiwan.}
\thanks{W.-Y. Chiu and C.-M. Chen are with the Department of Electrical Engineering, National Tsing Hua University,
Hsinchu 30013, Taiwan (e-mail: chiuweiyu@gmail.com).}
\thanks{H. Sun is with the Department of Engineering, Durham University, Durham, DH1 3LE, U.K.}
\thanks{\copyright2018 IEEE. Personal use of this material is permitted. Permission from IEEE must be obtained for all other uses, in any current or future media, including reprinting/republishing this material for advertising or promotional purposes, creating new collective works, for resale or redistribution to servers or lists, or reuse of any copyrighted component of this work in other works.}
\thanks{Digital Object Identifier 10.1109/JSYST.2018.2876345}
}

\maketitle

\begin{abstract}
This paper investigates smart home energy management in consideration of tradeoffs between residential privacy and energy costs.
A  multiobjective approach that minimizes energy costs and maximizes privacy protection is proposed.
The approach leads to a multiobjective optimization problem in which the two objectives are addressed in separate dimensions.
A hybrid algorithm that employs stochastic search for power scheduling of home appliances and
uses deterministic battery control is developed accordingly.
The proposed approach can avoid some drawbacks faced by conventional weighted-sum methods for multiobjective optimization:
the combination of objectives in different units, heuristic assignment of  weighting coefficients, and possible misrepresentation of user preference.
In contrast with existing studies on residential user privacy that assume limited controllability of appliances to facilitate algorithm development,
the approach addresses the use of flexible appliances in smart homes.
Simulations reveal that the proposed approach can maintain a reasonable energy cost while robustly preserving  user privacy at a sensible level;
its convergence rate is comparable to existing multiobjective evolutionary algorithms while the proposed approach yields a better level of convergence;
the proposed approach is scalable to a group of smart houses, achieving a superior peak-to-average ratio that is beneficial to the stability of the underlying power grid.
\end{abstract}


\IEEEpeerreviewmaketitle

\section{Introduction}
 With a real-time monitoring system\cite{7536160}, information transmission system\cite{7106508}, demand-side management system\cite{7042318}, and Internet of Things  technologies\cite{7721747,7805265}, smart grids are gradually replacing  traditional power grids as the new generation of power systems\cite{6819778,6803878,6762856}.
 One of the main features that separate the smart grid from its predecessor is its ability to effectively control power use to achieve energy conservation goals\cite{6712064}. Consequently, the expectation of smart grids also leads to a higher standard of security and privacy\cite{6194398,7828093,7360878,6949126}.

During  information collection and exchange in the smart grid, an attacker may steal or create fake power information to jeopardize the power system or obtain monetary rewards\cite{7588049,5054916}.
To secure the grid, an encryption or data obfuscation scheme is often used to protect data transmissions\cite{6506075,7134810,7571177}.
 Regardless of which scheme is employed, the utility grid
  ultimately requires access to the power consumption
 information  of residential users so that both power supply and demand can be balanced.
If data is not gathered and collected appropriately, then the life patterns (including any activities or even occupancy) of a user may be revealed\cite{7823349}, effectively raising privacy concerns.

There are a few ways to address smart grid privacy.
One popular way is to change normal power consumption patterns using various resources so that it is difficult to infer the true activities  of a particular residence.
For example, a water heater can be controlled in such a way that the smart meter reading shows a home is occupied at all times\cite{7050259};
energy harvesting and/or  storage units can be employed to cover the power usage\cite{6936925,6547840}.

A residential user must trade privacy for energy costs (or energy waste, power usage)\cite{7444814,6547840}.
Several studies have been conducted to find a balance between these two factors.
A weighted-sum method is the most common way to combine objectives pertaining to the privacy and cost into a single function\cite{7354489},
but limited controllability for home appliances is often assumed to facilitate the development of efficient algorithms.
For example, loads were modeled as random variables and dynamic programming was used in\cite{6876215,6847975,PRIVATUS}.
In~\cite{7501552}, cost-friendly differential privacy-preserving schemes were proposed in which energy consumption of appliances was given in advance and used as a system input.
In~\cite{7827910}, constant power consumption of appliances was assumed and a water-filling algorithm was employed to compute an optimal energy management policy.
In\cite{6522908}, unshiftable appliances were considered and  stochastic optimization via Monte Carlo simulation was proposed to control battery charging and discharging.

For trading residential privacy off against energy costs, there are two issues faced by most existing approaches.
First, appliances in smart houses often have certain flexibility in time of use and power of use, which is essential for smart functionality.
Second, combining objectives representing privacy and costs into a single objective using weighted-sum methods can be problematic\cite{marler2004survey,das1997closer,coit2004system,B_MOEA1}:
 it may not be meaningful to combine two quantities in different units and scales together, even after certain normalization;
a weighting coefficient is heuristically assigned, and no systematic way exists for such assignment;
 it is also unclear how to assign the coefficient that can precisely reflect a preference model of a residential user, i.e., precise quantification for the importance of privacy and cost.

In this paper, we consider various types of home appliances. These types are inflexible
and unshiftable appliances, flexible appliances, and shiftable appliances. We formulate a multiobjective optimization problem (MOP) that addresses a cost metric and privacy metric in different dimensions.
The privacy metric is defined using load variation. A hybrid algorithm consisting of stochastic and deterministic search mechanisms is  developed to solve the MOP.
After the MOP is solved, an energy management system based on batteries can then be established to schedule power consumption of appliances.

In the hybrid algorithm, the stochastic search determines the power scheduling of flexible and shiftable home appliances in response to power prices and a variance function of privacy, leading to a set of Pareto optimal solutions. Final scheduling for home appliances is determined by the solution that minimizes its distance to an ideal solution on the basis of an approximate Pareto front, i.e., the image of an approximate Pareto optimal set through the objective functions. By fixing the power consumption patterns of home appliances, the battery status is then determined using the following deterministic method: the  battery status is adjusted so that
current power consumption of the smart home is as close as possible to its previous power consumption. In this way, a smooth power consumption curve can be achieved, further refining residential  privacy.


The main contributions of this work are as follows. First, we formulate the privacy--cost problem in smart houses as an MOP. Objectives are addressed in different dimensions, which is different from conventional weighted-sum approaches that combine them into one single objective. The problem of different units in objectives, requirement of prior information for objective normalization, and heuristic assignment of weighting coefficients can be avoided.
Second, we address various types of appliances in smart homes, including time- and power-flexible appliances,
while existing studies on the privacy--cost problem often  assume limited controllability for appliances to develop efficient algorithms.
This bridges the gap between residential privacy of smart houses and the use of flexible home appliances.
Third, we propose a hybrid algorithm that searches for a set of Pareto optimal solutions to the MOP. The final tradeoff between the energy cost and privacy
is determined with the help of knowledge extracted from the whole solution set, bringing a broad perspective on optimality.
Finally, we validate the proposed algorithm through numerical analysis. Our simulations show that the proposed algorithm is more suitable to address user privacy than
 existing multiobjective evolutionary algorithms, with a comparable convergence rate and better level of convergence.

The remainder of this paper is organized as follows. Related work is briefly discussed in Section~\ref{sec_rel_work}.
 Section~\ref{sec_model} describes the system architecture, mathematical models of various home appliances, and energy storage device. The multiobjective approach is developed in Section~\ref{sec_MO}. Simulation results involving comparison among existing energy management methods are presented in Section~\ref{sec_sim}. Finally, Section~\ref{sec_conclusion} concludes this paper.

\section{Related Work}\label{sec_rel_work}

 Smart home privacy can be defined in many ways and relevant mechanisms for privacy preservation have been proposed\cite{privacy_review,privacy_review2}.
One of the most popular definitions is based on the load variance\cite{6847975,6876215,7354489,7827910,7050259}.
Perfect privacy is attained if the power consumption of a smart house measured from a smart meter remains constant.
In this case there is no way to infer any residential activities. Privacy preservation measured by the load variance is often realized by employing battery charging and discharging activities to change load profiles\cite{LV_battery1,LV_battery2}.

Another privacy measure is the use of mutual information that involves  statistics of residential power demand.
In\cite{6936925}, privacy was considered in a multiuser smart meter system in which an alternative energy source was involved. An explicit lower bound on a privacy--power function was derived.
In\cite{6308749}, a framework based on information theory and a hidden Markov model was proposed. A privacy--utility tradeoff problem with minimal assumptions was formulated.
In\cite{6547840}, mutual information was investigated in the presence of energy harvesting and storage units. Privacy was preserved by diversifying energy sources through energy harvesting.
When mutual information is employed for privacy measure, battery use is still the dominant way to improve user privacy\cite{MI_battery1,MI_battery2,MI_battery3}.

In addition to the use of load variance and mutual information for privacy measure, some studies have considered differential privacy\cite{DP_1,DP_2,7501552}.
No matter which privacy measures have been used, the aforementioned studies have adjusted the battery status to improve user privacy without considering possible adjustment on flexible or shiftable appliances. Limited controllability of appliances has been assumed to facilitate the development of  efficient algorithms.

In practice, smart houses often involve a few appliances that are flexible in time of use or power of use\cite{SHIRAZI201540,7095598,MA2016320,Jordi}.
Those flexible appliances can  be controlled to minimize energy costs or reduce peak load.
In\cite{7948743}, demand side management using evolutionary algorithms was proposed for scheduling appliances of residential users.
In\cite{rastegar2016home},
an optimization problem in which customer energy and reliability costs were minimized was investigated.
In\cite{6895131},
user's convenience level, thermal comfort level, and energy cost were mixed as a ratio, leading to a multiobjective approach. A weighting coefficient representing their degrees of importance
was prescribed heuristically.

Aforementioned papers in which flexible appliances are controlled for energy management  have not considered user privacy, jeopardizing further participation of home residents in grid activities.
In\cite{7202880}, a decentralized framework for load management was developed and user privacy was claimed to be protected, but no explicit measure for privacy protection was presented and optimized. There seems to be a gap between residential privacy of smart houses and the use of flexible home appliances, which is addressed in this study.

\section{System Models}\label{sec_model}

This section presents the model of a smart home connected to an energy storage device and the utility grid.
In the smart home, the energy storage device and utility grid provide energy to home appliances; a control unit schedules power consumption of appliances
in response to time-varying market prices.

\subsection{Home Appliances}
Home appliances are classified  into three categories: inflexible and unshiftable, flexible, and shiftable appliances.
For inflexible and unshiftable appliances, time of use and the associated power consumption are fixed.
Examples of this category are as follows:
    a refrigerator that is turned on for a duration of 24 hours possesses a power consumption curve that is mostly fixed once the cooling temperature has been set; cooking appliances, such as ovens and electric pots, are often used in lunch/dinner time slots with fixed power consumption;
  lights are turned on during specific time slots, consuming  a fixed amount of power; televisions or computers
  are mostly used after work with fixed power consumption.
 Let $\mathcal{A_{IU}}$ denote the index set of all inflexible and unshiftable home appliances. The power consumption in time slot $h$ is denoted by $P_{a}^{h}$ kW, where $a \in \mathcal{A_{IU}}$.

For flexible home appliances, the associated power consumption can be controlled and time of use can be shifted.
One example of this category is an air conditioner.
Let $ \mathcal{A_{F}}$ denote the index set of flexible home appliances. The power consumption of appliance $b\in \mathcal{A_{F}}$ in time slot $h$ is denoted by $P_{b}^{h}$ kW.
The value of  $P_{b}^{h}$ is affected by the starting time slot of use $s_{b}$, the ending time slot $e_{b}$, the minimum power consumption $P_{b}^{\min}$, and the maximum power consumption $P_{b}^{\max}$. In general, we have
\begin{equation}\label{eq_P_AF}
\begin{split}
\left\{
  \begin{array}{ll}
P_{b}^{\min}< P_{b}^{h} \leq P_{b}^{\max},  \hbox{ if }  h\in  \{s_{b},s_{b}+1,...,e_{b}-1,e_{b}\} ;    \\
P_{b}^{h}=0 , \quad \hbox{ otherwise }
  \end{array}
\right.
\end{split}
\end{equation}
where $e_{b}>s_{b}$.
It is worth mentioning that this category includes power-flexible appliances, which can be modeled by specifying
the starting and ending time slots for a particular appliance in advance.
In this case, power consumption $P_{b}^{h}$ is a decision variable that can be adjusted and determined from optimization.

For shiftable home appliances, power consumption  cannot be controlled, but time of use can be shifted as desired.  A washing machine, for example, can be turned on in any time slot with a fixed working duration.  Let $\mathcal{A_{S}}$ denote the index set of shiftable appliances. The power consumption in time slot $h$ is denoted by $P_{c}^{h}$ kW, where  $c\in \mathcal{A_{S}}$.
 Let $h_{c}$ be the set of chosen time slots with $|h_{c}|$ representing the number of time slots required to finish the work.
 Given the starting time slot $s_{c}$ and ending time slot $e_{c}$
 for appliance $c$ ($s_{c}<e_{c}$), the power consumption of appliance $c$ satisfies
\begin{equation}\label{eq_P_AS}
\begin{split}
\left\{
  \begin{array}{ll}
P_{c}^{h}(h_{c})>0, & \hbox{ if }  h \in h_{c}; \\
P_{c}^{h}(h_{c})=0 , & \hbox{ otherwise.}
  \end{array}
\right.
\end{split}
\end{equation}

The minimum power consumption for flexible appliances and starting time for shiftable appliances
can affect the user comfort. It is preferred that power consumption is close to normal consumption
and the starting time is close to the request time\cite{Heu}.
In this study, the user comfort can be addressed by
prescribing the minimum power consumption and/or starting time of appliances, imposing an additional constraint.
The method of using constraints to address user comfort has been widely adopted\cite{WANG2016613,6365816,Jordi}. In our scheme, the constraint related to user comfort would specify the lower bound for the minimum power consumption and/or upper bound for the starting time. When the constraint is satisfied, a level of user comfort is guaranteed.
It is worth mentioning that this method is simplistic. For example, user thermal comfort can depend on several factors including external climate and occupancy level. In this case more complex models for user comfort should be considered.

\subsection{Energy Storage Device}

In this study, a battery serves as an energy storage device. The status of the battery in time slot $h$ is denoted by $B(h)$ kWh, satisfying
\begin{equation}\label{eq_Battery_bound}
\begin{split}
B^{\min}\leq B(h)\leq B^{\max}
\end{split}
\end{equation}
where $B^{\min}$ and $B^{\max}$ are the minimum battery level and maximum capacity, respectively.
The quantity $B(h)$ is mainly affected by  charging and discharging activities, which are modeled by $S(h)$ kW.
We assume that positive power means charging and  negative power  means discharging.
The power charging/discharging  profile in time slot $h$ can be further expressed as $S(h)=S^{+}(h)-S^{-}(h)$, where
\begin{equation}\label{eq_S_bound}
\begin{split}
 S^{+}(h){ }={ }&\mathop{\max}\{S(h),0\}  \quad  \mbox{and}  \\
S^{-}(h){ }={ }&-\mathop{\min}\{S(h),0\}.
\end{split}
\end{equation}

 In the process of power conversion,
some power losses may occur. This can be modeled by
using the charging efficiency  $\beta^{+}$  and  discharging efficiency $\beta^{-}$,
where $0<\beta^{+}\leq 1$  and  $\beta^{-}\geq 1$.
For the charging mode, the utility grid supplies power $S^+(h)$ but only $\beta^{+}S^+(h)$ is stored to the battery; for the discharging mode, the battery discharges power  $\beta^{-}S^{-}(h)$ but only $S^{-}(h)$ is consumed by home appliances.
 In addition, the battery status can decrease over time with a leakage rate $\alpha$, where $0<\alpha\leq 1$.
The battery profile can thus be expressed as\cite{6297498,7862931,6805662}
\begin{equation}\label{eq_battery_state}
\begin{split}
B(h+1)=\alpha B(h)+\beta(h)S(h)
\end{split}
\end{equation}
where
\begin{equation}\label{eq_Beta}
\begin{split}
{
\beta(h)=
\left\{
  \begin{array}{ll}
\beta^{+},& \hbox{ if } S^{+}(h)>0;\\
\beta^{-},& \hbox{ if } S^{-}(h)>0.
  \end{array}
\right.
}
\end{split}
\end{equation}

According to~(\ref{eq_battery_state}),
the power profile can be calculated as
\begin{equation}\label{eq_S_operate}
\begin{split}
S(h)=
\left\{
  \begin{array}{ll}
\frac{B(h+1)-\alpha B(h)}{\beta^{+}}, & \hbox{ if } S^{+}(h)>0;\\
\frac{B(h+1)-\alpha B(h)}{\beta^{-}}, & \hbox{ if } S^{-}(h)>0.
  \end{array}
\right.
\end{split}
\end{equation}
Let $S^{\max}$ be  the maximum power that can be provided from the grid to the battery or discharged from the battery to home appliances in each time slot.  We have $\beta \left\vert S(h) \right\vert \leq S^{\max}$.
The battery's capacity range in time slot $h+1$ is then bounded as
\begin{equation}\label{eq_Battery_next_status}
\begin{split}
\alpha B(h)-S^{\max}\leq B(h+1)\leq \alpha B(h)+S^{\max}.
\end{split}
\end{equation}

Finally, the power supply form the battery to   home appliances satisfies
\begin{equation}\label{eq_S_and_PHA}
\begin{split}
S(h)\leq P_{HA}^{h}
\end{split}
\end{equation}
where
\begin{equation}\label{eq_PHA}
\begin{split}
P_{HA}^{h}=\sum_{a \in \mathcal{A_{IU}}}P_{a}^{h}+\sum_{b \in \mathcal{A_{F}}}P_{b}^{h}+\sum_{c \in \mathcal{A_{S}}}P_{c}^{h}
\end{split}
\end{equation}
is the total power consumption of all kinds of home appliances.
When the equality in~(\ref{eq_S_and_PHA}) holds true,
the power from the battery is enough to sustain all the home appliances. In this case no additional power from the grid is needed.

\subsection{Electricity Price}

Dynamic pricing schemes have been extensively adopted to adjust power loads. Some popular pricing schemes include the day-ahead pricing and real-time pricing\cite{7353201, 6985863, 6730904,6462005}.
Although a real-time pricing scheme is promising for the minimization of energy costs
by controlling the power usage of flexible appliances in response to market price variations,  online optimization is needed, leading to greater computational complexity.
 This paper uses  a day-ahead pricing  scheme for a few reasons\cite{NordPool,AEP_ENERGY,IESO}.
 First, a day-ahead market has been widely adopted: it serves the main way of  trading power  and allows the majority of system load to be committed in some regions. Second, it can improve operational certainty for system operators and financial certainty for dispatchable resources.
 Third, it can avoid price volatility due to abrupt changes in supply and demand in a real-time market.
In addition, day-ahead pricing schemes have been considered in a number of recent studies on related topics (see, for example, \cite{8012490,7457198,8053827} and\cite{7305766}), showing its popularity among other pricing schemes.

Let
\begin{equation}\label{eq_total_power}
P^{h}=P_{HA}^{h}+S(h).
\end{equation}
The power consumption cost in a residential home can be calculated as
\begin{equation}\label{eq_obj_cost}
\begin{split}
F_{Cost}=\sum_{h\in \mathcal{H}}(P^{h}\times \Delta h)\times\lambda^{h}
\end{split}
\end{equation}
where $\mathcal{H}$ represents the index set of time slots and
$\lambda^{h}$ represents the market price in time slot $h$.
For the time intervals of 1 hour, 30 minutes, and 15 minutes,
we have $(\Delta h, |\mathcal{H}|)=(1,24)$, $(\Delta h, |\mathcal{H}|)=(0.5,48)$, and
$(\Delta h, |\mathcal{H}|)=(0.25,96)$, respectively.

\section{User-Centric Multiobjective Approach}\label{sec_MO}

A residential user may desire to minimize the energy cost while keeping his/her power consumption profile confidential.
To avoid leaking user information to the utility grid,  one can use the energy storage device to render the power usage curve as flat as possible, disguising the actual power consumption patterns. However, a flat pattern can result in high energy costs when market prices vary over time.
This section develops a user-centric multiobjective approach that considers the energy consumption cost and   privacy preservation simultaneously.

By suitably scheduling power consumption patterns,  the energy costs can be lowered.
 For appliances indexed by $\mathcal{A_{F}}$ and $\mathcal{A_{S}}$, we may shift the time of use or reduce the amount of power usage when the market price is high.
  This practice can be achieved by solving
\begin{equation}\label{eq_opt_cost}
\begin{split}
\mathop{\min}_{\substack{P_{b}^{h}, h_{c}, B(h)\\ h\in \mathcal{H},b\in \mathcal{A_{F}}\\c\in \mathcal{A_{S}}}}F_{Cost}
\end{split}
\end{equation}
where
there are $\left\vert \mathcal{H}\right\vert \times \left\vert \mathcal{A_{F}}\right\vert$ for  $P_{b}^{h}$-type decision variables,  $(e_c-s_c+1) \times \left\vert \mathcal{A_{S}}\right\vert$ for $h_{c}$-type
decision variables,  and $\left\vert \mathcal{H}\right\vert$  for $B(h)$-type decision variables.

While reshaping the power consumption pattern in response to market prices  could be financially beneficial,
 the resulting pattern may reveal the activities of a residential user.
 In general, the  residential privacy can be measured by a variance function\cite{7354489,7827910,7050259}:
 \begin{equation}\label{eq_obj_privacy}
\begin{split}
F_{Privacy}=\frac{1}{\left\vert \mathcal{H}\right\vert}\sum_{h\in \mathcal{H}}(P^{h})^2-(\frac{1}{\left\vert \mathcal{H}\right\vert}\sum_{h\in \mathcal{H}}P^{h})^2
\end{split}
\end{equation}
where $P^{h}$ is defined in~(\ref{eq_total_power}).
A smaller value of $F_{Privacy}$ implies
a more flat power consumption profile, yielding a better level of privacy.
 Ideal privacy preservation occurs when  $F_{Privacy}=0$, i.e., the power consumption is time-invariant.
Privacy can thus be preserved by solving the minimization problem
\begin{equation}\label{eq_opt_privacy}
\begin{split}
\mathop{\min}_{\substack{P_{b}^{h}, h_{c}, B(h)\\ h\in \mathcal{H},b\in \mathcal{A_{F}}\\c\in \mathcal{A_{S}}}}F_{Privacy}.
\end{split}
\end{equation}

 Referring to~(\ref{eq_opt_cost}) and~(\ref{eq_opt_privacy}),
we solve the following MOP to consider both the energy cost and privacy:
\begin{equation}\label{eq_MOP_cost0}
\begin{split}
&\mathop{\min}_{\substack{P_{b}^{h}, h_{c}, B(h)\\ h\in \mathcal{H},b\in \mathcal{A_{F}}\\c\in \mathcal{A_{S}}}}\text{ }F_{Cost}\\
&\mathop{\min}_{\substack{P_{b}^{h}, h_{c}, B(h)\\ h\in \mathcal{H},b\in \mathcal{A_{F}}\\c\in \mathcal{A_{S}}}}\text{ }F_{Privacy}\\
&\text{subject to}~(\ref{eq_P_AF}),~(\ref{eq_Battery_bound}),~(\ref{eq_Battery_next_status}), \text{ and}~(\ref{eq_S_and_PHA}).
\end{split}
\end{equation}
The MOP has
\begin{equation}\label{eq_num_dec}
   \left\vert \mathcal{H}\right\vert \times   ( \left\vert \mathcal{A_{F}}\right\vert +1 )    +  (e_c-s_c+1) \times \left\vert \mathcal{A_{S}}\right\vert
\end{equation}
decision variables.

Since the objectives are conflicting, a global optimal solution does not exist and thus Pareto optimality is adopted\cite{B1,J10}.
At first, it seems that~(\ref{eq_MOP_cost0}) is a typical MOP that can be solved by existing multiobjective evolutionary algorithms.
In practice, however, the battery constraint in~(\ref{eq_Battery_next_status}) can bring difficulty to the solving process when the decision variable space is searched directly for feasible candidate solutions. This is because
  $B(h)$ in each time slot serving as a decision variable introduces $|\mathcal{H}|$ dynamical constraints,
  which are hard to satisfy by any generic random search.
To address this difficulty, we replace $B(h)$ with the power
charging/discharging profile $S(h)$ and propose a  hybrid algorithm consisting of both stochastic and deterministic search mechanisms. Once $S(h)$  has been obtained, the battery status can then be determined  by~(\ref{eq_battery_state}).\footnote{The basic framework of the hybrid algorithm is motivated by artificial immune systems and best-effort algorithms.
For stochastic exploration, the hybrid algorithm mimics the evolutionary strategy
of artificial immune systems in which antibodies (or candidate solutions) evolve to combat antigens\cite{7050260,B2}.
For deterministic exploitation, the hybrid algorithm employs the information about the difference of power consumption  in two consecutive time slots to smooth the power usage, which is similar to best-effort moderation algorithms\cite{LV_1,MI_battery1}.}

To apply our hybrid algorithm, we
 define the new objective functions as
\begin{equation}\label{eq_newcost}
\begin{split}
F_{C}=\sum_{h\in \mathcal{H}}(P_{HA}^{h}\times \Delta h)\times\lambda^{h}
\end{split}
\end{equation}
and
\begin{equation}\label{eq_newprivacy}
\begin{split}
F_{P}=\frac{1}{\left\vert \mathcal{H}\right\vert}\sum_{h\in \mathcal{H}}(P_{HA}^{h})^2-(\frac{1}{\left\vert \mathcal{H}\right\vert}\sum_{h\in \mathcal{H}}P_{HA}^{h})^2.
\end{split}
\end{equation}
The new MOP is then expressed as
\begin{equation}\label{eq_MOP_cost}
\begin{split}
&\mathop{\min}_{\substack{P_{b}^{h}, h_{c}\\ h\in \mathcal{H},b\in \mathcal{A_{F}}\\c\in \mathcal{A_{S}}}}\text{ }F_{C}\\
&\mathop{\min}_{\substack{P_{b}^{h}, h_{c}\\ h\in \mathcal{H},b\in \mathcal{A_{F}}\\c\in \mathcal{A_{S}}}}\text{ }F_{P}\\
&\text{subject to }~(\ref{eq_P_AF}).
\end{split}
\end{equation}
The problem in~(\ref{eq_MOP_cost}) is different from the original problem in~(\ref{eq_MOP_cost0}): it considers power scheduling without the use of batteries and therefore
the battery constraints described by (\ref{eq_Battery_bound}), (\ref{eq_Battery_next_status}),  and~(\ref{eq_S_and_PHA}) are excluded.
The idea for solving~(\ref{eq_MOP_cost0}) is to stochastically explore various power scheduling profiles without the use of batteries by solving~(\ref{eq_MOP_cost}) (no battery constraints involved), and then compensate it by deterministically smoothing the power usage in two consecutive time slots with the help of batteries (battery constraints involved).

Algorithm 1 presents the pseudocode of our hybrid algorithm, addressing not only the solution to~(\ref{eq_MOP_cost}) but also the smoothing process for power usage.
Because of the hybrid procedure (stochastic exploration followed by deterministic exploitation), our solution method for~(\ref{eq_MOP_cost0}) is considered as sub-optimal.
The concept of domination in Algorithm~1 is Pareto domination with respect to~(\ref{eq_MOP_cost}).
 Let $t_{c}$ be the algorithm counter. The maximum number of iterations is set to be $t_{\max}$.
The antibody population  is represented by $\mathcal{X}(t_{c})$ and its size is denoted as $|\mathcal{X}(t_{c})|$.
Each antibody $\bm{x}\in \mathcal{X}(t_{c})$ consists of decision variables $P_{b}^{h}$ and $h_{c}$. The nominal population  size  and maximum population size are denoted by $N_{\text{nom}}$ and $N_{\max}$, respectively.

\begin{algorithm}
\caption{Proposed Hybrid Algorithm}
\label{alg:code}
\begin{algorithmic}
\Require Nominal size of population $N_{\text{nom}}$, maximum  population size $N_{\text{max}}$, and maximum iteration times $t_{\text{max}}$;
    Day-ahead price $\lambda^h$, $P_{a}^{h}$, $P_{b}^{\min}$, $P_{b}^{\max}$, $s_{b}$,
  $e_{b}$, $P_{c}$, $s_{c}$, $e_{c}$, $t_{P_{c}}$, and battery parameters $\beta^{+}$, $\beta^{-}$,
 $\alpha$, $S^{\max}$, $B^{\min}$, and $B^{\max}$.
\Ensure Power scheduling represented by $\bm{x}^{*}$ and $S(h), h\in \mathcal{H}$.
\State Step 1) Initialize $\mathcal{X}(0)$.
\State Step 2) Remove dominated antibodies  from $\mathcal{X}(0)$; let $t_{c}=0$.
\State  \textbf{while} $t_c \leq t_{\text{max}}$ \textbf{do}
\State Step 3) Clone antibodies $\bm{x}$.
\State Step 4) Apply gene operations to $P_{b}^h$  over
$[P_{b}^{\min},P_{b}^{\max}]$ for flexible home appliances; generate
working hours $h_{c}$ for shiftable home appliances.
\State Step 5) Remove dominated antibodies from $\mathcal{X}(t_c)$.
\State Step 6) Maintain a manageable size of $\mathcal{X}(t_c)$;
 let $\mathcal{X}(t_c+1)=\mathcal{X}(t_c)$ and $t_c=t_c+1$.
 \State  \textbf{end while}
\State Step 7) Output approximate Pareto set  $\mathcal{X}(t_{\max})$.
\State Step 8) Select the final solution $\bm{x}^{*} \in \mathcal{X}(t_{\max})$.
\State Step 9) Employ the charging/discharging profile $S(h),h \in \mathcal{H}$  to smooth the total power consumption
on the basis of the power consumption change of home appliances.

\end{algorithmic}
\end{algorithm}

  Algorithm~1 is detailed as follows. Antibody population is initialized in  Steps 1) and 2).
In Steps 3) and 4), antibodies evolve over iterations for quality improvement.
 The clonal rate is $\llcorner N_{\text{max}}/|\mathcal{X}(t_{c})| \lrcorner$, where $\llcorner \cdot \lrcorner$ represents the floor function.
In Steps 5) and 6),  dominated antibodies are removed from  $\mathcal{X}(t_{\max})$ and the population set is updated.
The size of $\mathcal{X}(t_c)$ is considered as manageable if $| \mathcal{X}(t_c)|\leq N_{\text{nom}}$ holds true.
Steps  3)--6) are repeated until  the maximum iteration is reached.
Step 7) outputs the approximate Pareto set.  In Step 8), the minimum Manhattan distance approach in\cite{7465803} is applied to select the final solution  $\bm{x}^{*}$.
First,
the maximum spread of the approximate Pareto front in each dimension is evaluated:
\begin{equation}\label{alg:MMD_L}
\begin{split}
L_{C}{ }={ } & \underset{ {\bm{x}} \in \mathcal{X}(t_{\max})}{\text{max}} F_{C}(\bm{x})-\underset{ {\bm{x}} \in \mathcal{X}(t_{\max})}{\text{min}} F_{C}(\bm{x}) \mbox{ and }\\
L_{P}{ }={ } &  \underset{ {\bm{x}}\in \mathcal{X}(t_{\max})}{\text{max}} F_{P}(\bm{x})-\underset{ {\bm{x}} \in \mathcal{X}(t_{\max})}{\text{min}} F_{P}(\bm{x}).
\end{split}
\end{equation}
Second, define
\begin{equation*}\label{alg:MMD_point}
\bm{y}(\bm{x})=
\left[
  \begin{array}{cc}
     \frac{ F_{C}(\bm{x})}{L_{C}} & \frac{ F_{P}(\bm{x})  }{L_{P}} \\
  \end{array}
\right]^T
\end{equation*}
and the normalized ideal vector as
\begin{equation*}\label{alg:MMD_ideal}
\bm{y}^*_{idl}=
\left[
  \begin{array}{cc}
     \frac{\underset{ {\bm{x}} \in \mathcal{X}(t_{\max}) }{\text{min}} F_{C}(\bm{x})}{L_{C}} & \frac{\underset{ {\bm{x}} \in \mathcal{X}(t_{\max})}{\text{min}} F_{P}(\bm{x})  }{L_{P}} \\
  \end{array}
\right]^T.
\end{equation*}
Finally, the solution $\bm{x}^{*}$ is selected by
\begin{equation}\label{eq_MMD}
\bm{x}^{*}=  \arg \underset{{\bm{x}\in\mathcal{X}(t_{\max})}}{\text{min}} \; || \bm{y}(\bm{x}) - \bm{y}^*_{idl}||
\end{equation}
where $|| \cdot||$ represents the Manhattan norm.

The procedure before Step 9) is used to determine the power consumption of home appliances.
To complete power scheduling in a smart home, a rule is required to control the battery status.
To this end, Step 9) employs the power charging/discharging profiles $S(h),h \in \mathcal{H}$
to determine the battery status $B(h)$ in consideration of user privacy. Let
 \begin{equation}\label{eq_delta_P}
    \Delta P_{HA}^{h}=P_{HA}^{h}-P_{HA}^{h-1}
 \end{equation}
 denote the power consumption change of home appliances in two consecutive time slots.
 We design the profile $S(h)$ as
\begin{equation}\label{alg_operate_S}
\begin{split}
{\scriptsize
S(h)=
\left\{
  \begin{array}{ll}
0, & \hbox{if } \left\vert \Delta P_{HA}^{h}\right\vert \leq \epsilon\\
\mathop{\min} \{S^{\max},- \Delta P_{HA}^{h} ,B^{\max}-\alpha B(h)\}, & \hbox{if } \Delta P_{HA}^{h}  < -\epsilon \\
-\mathop{\min} \{S^{\max}, \Delta P_{HA}^{h},\alpha B(h)-B^{\min}\}, & \hbox{if } \Delta P_{HA}^{h} > \epsilon
  \end{array}
\right.
}
\end{split}
\end{equation}
where $\epsilon$ is a small positive constant.

The charging/discharging operation
in~(\ref{alg_operate_S}) can effectively disguise residential activities and
satisfy the dynamical constraints induced by the battery. This is the deterministic step integrated into the hybrid algorithm.
When the power consumption of home appliances in two consecutive time slots remains approximately constant (i.e., $ |\Delta P_{HA}^{h} |< \epsilon $),
the battery is neither charged nor discharged (i.e., $S(h)=0$).
When the power consumption of home appliances decreases (i.e., $ \Delta P_{HA}^{h} < - \epsilon $),
the battery is charged (i.e., $S(h)>0$) to cover the difference  $- \Delta P_{HA}^{h}$ but the charging rate should be lower than the maximum rate $S^{\max}$.
The energy provided to the battery should be upper bounded by the remaining capacity $B^{\max}-\alpha B(h)$.
When the power consumption of home appliances increases (i.e., $ \Delta P_{HA}^{h} > \epsilon $),
the battery is discharged (i.e., $S(h)<0$) to cover the difference $\Delta P_{HA}^{h}$; however, the discharging rate should be lower than the maximum rate $S^{\max}$.
The energy provided by the battery should be upper bounded by the remaining energy $\alpha B(h)-B^{\min}$.
In the event that  the battery status is lower than $B^{\min}$ (i.e., $\alpha B(h)-B^{\min}<0$), the battery is charged.

In general, employing deterministic steps for exploitation in hybrid algorithms can increase the execution time.
This is because stochastic exploration is first performed and then followed by deterministic exploitation, with overlapping search of certain portion of the decision space.
However, this is not the case for the proposed hybrid algorithm in which the deterministic step does not overlap stochastic exploration  in decision space
but replaces partial stochastic exploration.
The computational complexity of our hybrid algorithm is related to the size of the problem, and the problem size is related to the number of decision variables involved.
As shown in~(\ref{eq_num_dec}), this number is mainly determined by the chosen time interval and the number of appliances in smart houses, which is generally manageable and does not increase over time.
In the proposed scheme, the computational complexity, particularly the execution time, would not be a big concern.
The reason is that power scheduling of appliances is optimized on the basis of day-ahead pricing signals, which means that only offline optimization is required.

\begin{figure*}
  \centering
    \includegraphics[width=\textwidth,height=7cm]{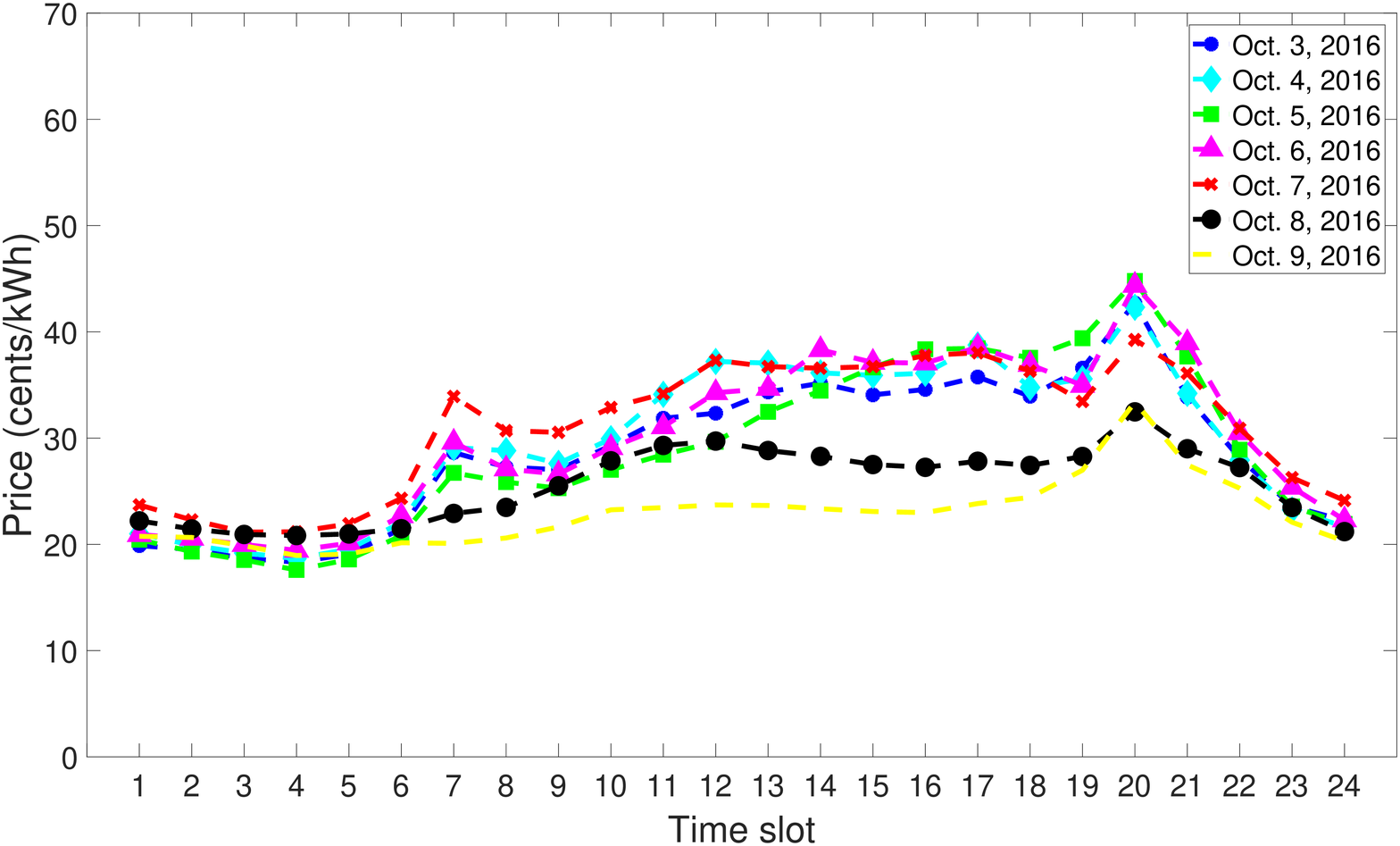}\\
  \caption{Pricing schemes of one week.}\label{fig_price}
\end{figure*}
\begin{figure*}
  \centering
    \includegraphics[width=\textwidth,height=7cm]{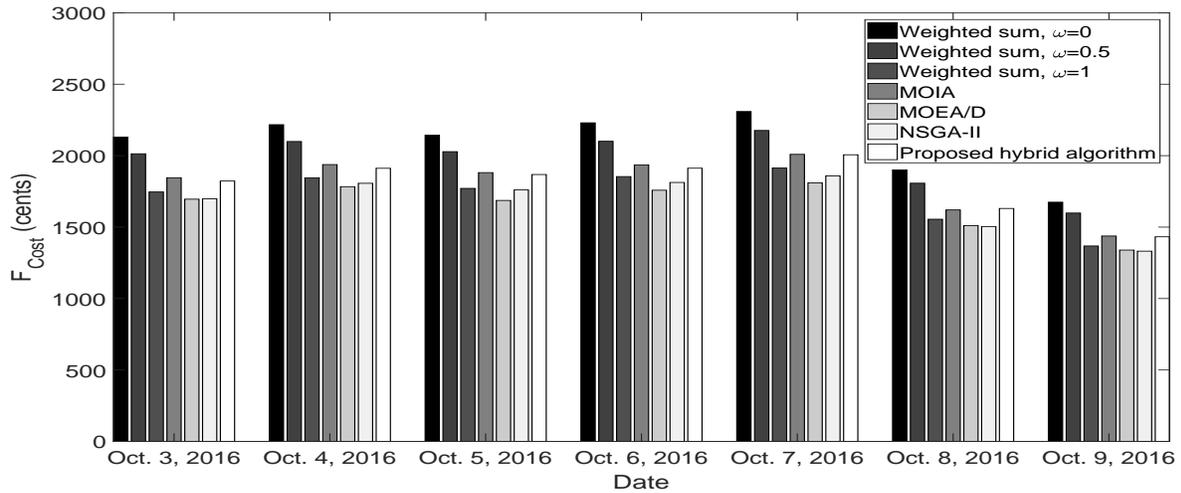}\\
  \caption{Energy costs of different power scheduling methods  in one week.}\label{fig_costweek}
\end{figure*}
\begin{figure*}
  \centering
    \includegraphics[width=\textwidth,height=7cm]{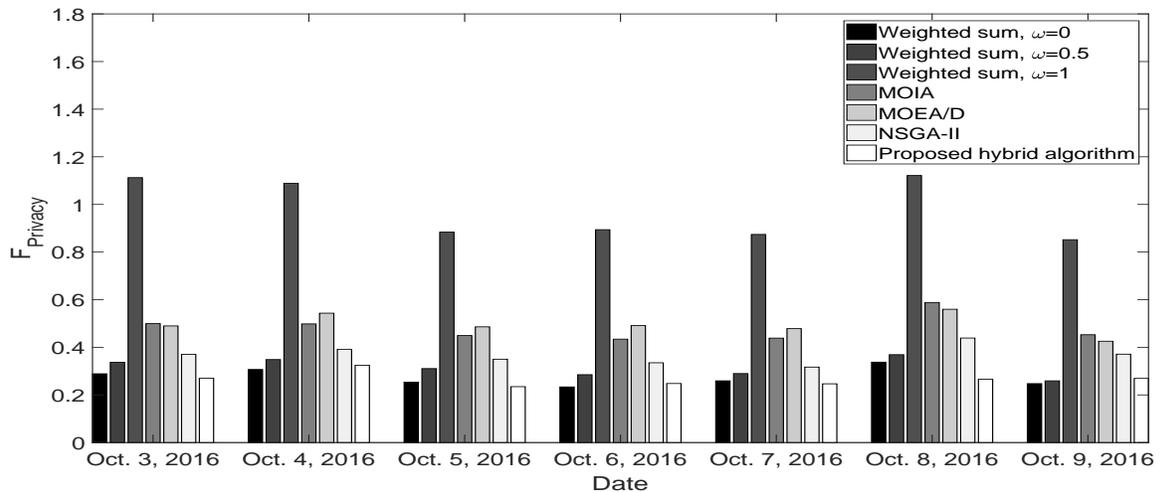}\\
  \caption{Privacy preservation of different power scheduling methods in one week. A smaller value of $F_{Privacy}$ means better privacy.}\label{fig_privacyweek}
\end{figure*}

\begin{table}
\centering
\caption{Parameters of Home Appliances}
\label{Loads}
\resizebox{\columnwidth}{!}{
\begin{tabular}{|c|c|c|}
\hline
Inflexible and Unshiftable Home Appliances & \multicolumn{2}{c|}{Power Consumption}  \\ \hline
$P_{a_1}^h$                  & \multicolumn{2}{c|}{0.015 kW  }                 \\ \hline
$P_{a_2}^h$                  & \multicolumn{2}{c|}{0.15 kW  }               \\ \hline
$P_{a_3}^h$                  & \multicolumn{2}{c|}{0.02 kW   }             \\ \hline
$P_{a_4}^h$                  & \multicolumn{2}{c|}{0.15 kW   }          \\ \hline
$P_{a_5}^h$                  & \multicolumn{2}{c|}{0.9 kW  }          \\ \hline
$P_{a_6}^h$                  & \multicolumn{2}{c|}{1.3 kW  }            \\ \hline
$P_{a_7}^h$                  & \multicolumn{2}{c|}{0.2 kW   }           \\ \hline
$P_{a_8}^h$                  & \multicolumn{2}{c|}{0.1 kW }          \\ \hline
$P_{a_9}^h$                  & \multicolumn{2}{c|}{0.05 kW  }           \\ \hline
$P_{a_{10}}^h$                  & \multicolumn{2}{c|}{1.5 kW  }              \\ \hline
$P_{a_{11}}^h$                 & \multicolumn{2}{c|}{1.4 kW   }            \\ \hline
$P_{a_{12}}^h$                 & \multicolumn{2}{c|}{0.2 kW }            \\ \hline
$P_{a_{13}}^h$                 & \multicolumn{2}{c|}{0.8 kW  }           \\ \hline
$P_{a_{14}}^h$                 & \multicolumn{2}{c|}{0.5 kW  }            \\ \hline
\hline
Flexible Home Appliance &Power Consumption & Parameters \\ \hline
$P_{b}^h$         & $P_{b}^{\min}=1$ kW, $P_{b}^{\max}=3$ kW & $s_{b}=1, e_{b}=24$\\ \hline
\hline
Shiftible Home Appliance &Power Consumption & Parameters \\ \hline
\multirow{2}{*}{$P_{c}^h$}         &\multirow{2}{*}{ $1$ kW} & $s_{c}=$ rand$(10,13)$,\\
&& $e_{c}=s_{c}+7, t_{P_{c}}=1$\\ \hline
\hline
Battery  & \multicolumn{2}{c|}{ Parameters}  \\ \hline
$\alpha$                  & \multicolumn{2}{c|}{$\sqrt[24]{0.9}$  }                 \\ \hline
$\beta^{-}$                  & \multicolumn{2}{c|}{1.1}               \\ \hline
$\beta^{+}$                  & \multicolumn{2}{c|}{0.9}               \\ \hline
$B^{\max}$                  & \multicolumn{2}{c|}{4 kWh}               \\ \hline
$S^{\max}$                  & \multicolumn{2}{c|}{0.5 kW}               \\ \hline
$B^{\min}$                  & \multicolumn{2}{c|}{1 kWh}               \\ \hline
\end{tabular}}
\end{table}

\section{Simulation Results}\label{sec_sim}
Power scheduling in a smart home over the course of a week was examined in this section.
The smart home was equipped with fourteen inflexible and unshiftable home appliances, one air conditioner unit (flexible home appliance), and one washing machine (shiftable home appliance). A lithium-ion battery with parameters from\cite{6297498} was used as an energy storage device.
The time interval of one hour was chosen, which is the setting adopted by current demand response programs in practical use.
The smart home was assumed to have enough resources to perform associated optimization.
Table~\ref{Loads} presents the parameter settings for  home appliances. The starting time slot $s_{c}= \text{rand}(10,13)$ for appliance $c$ means that $s_{c}$ is a discrete random variable over $\{10,11,12,13\}$.

Parameters of the proposed hybrid algorithm were set as follows: $N_{nom} = 50, N_{max} = 1000$, and $t_{max} = 2000$.
To confirm its effectiveness, we compared the proposed approach with the weighted-sum method in\cite{7354489},
 as well as benchmark evolutionary algorithms MOIA\cite{7050260}, NSGA-II\cite{6562791}, and MOEA/D\cite{4358754}.
The weighted-sum method and MOEA/D need prior information to normalize objective functions before optimization;
in our simulations, objectives
\begin{equation}\label{eq_normal_MOEAD}
  F_{Cost}:=F_{Cost}/2400   \mbox{ and } F_{Privacy}:=F_{Privacy}/1.4
\end{equation}
were normalized on the basis of their maximum values (prior information).
For the weighted-sum method, a weighting coefficient $\omega$ was used to balance a tradeoff between the energy costs and privacy, leading to single-objective optimization. When  $\omega=0$, only privacy was focused; when $\omega=1$, only energy cost was focused; when  $\omega=0.5$, both the energy cost and privacy were addressed.
 All simulations were performed using a desktop with Intel i7-6700, 3.40 GHz CPU, and 8.0 GB RAM.
 Details of theses methods can be found in the appendix.

\subsection{Case Study Using Real Pricing Signals}

Fig.~\ref{fig_price} shows the day-ahead pricing signals from PJM in 3--9 Oct. 2016\cite{Web2}.
Simulations results are presented  in
Fig.~\ref{fig_costweek}, Fig.~\ref{fig_privacyweek}, and Table~\ref{comparison_cost}.
The weighted-sum method with $\omega=1$
attained an excellent level of cost saving but
 yielded the worst level of privacy preservation, which can be undesirable to residential users;
 the weighted-sum methods with $\omega=0.5$ and $\omega=0$
were outperformed by our approach in terms of both cost saving and privacy preservation.

In contrast with weighted-sum methods, algorithms MOIA, MOEA/D, and NSGA-II  evaluated objectives in separate dimensions.
Compared with our hybrid algorithm,  a maximum 7.87 percent in cost saving was achieved by MOEA/D, but user privacy was degraded by 88 percent;
NSGA-II found a better balance than MOEA/D with 6.5 percent in cost saving and 39 percent in privacy degradation; MOIA yielded worst performance because it did not reduce energy costs but degraded the user privacy by approximately 82 percent on average.
For MOIA, MOEA/D, and NSGA-II, user privacy was severely compromised on Oct. 8, 2016.

\begin{table}
\centering
\caption{Comparison of Various Power Scheduling Methods}
\label{comparison_cost}
\resizebox{\columnwidth}{!}{
\begin{tabular}{|c|c|c|c|c|c|c|}
\hline
\multicolumn{7}{|c|}{Cost Increase of Residential Users With Respect to Proposed Hybrid Algorithm}\\ \hline
{Date}&$\omega =0$&$\omega =0.5$&$\omega =1$&MOIA&MOEA/D &NSGA-II\\ \hline
Oct. 3, 2016& 16.91\% &10.47\% &-4.19\% &1.27\% &-6.96\% &-6.85\%\\ \hline
Oct. 4, 2016& 15.96\% &9.8\% &-3.47\% &1.36\% &-6.73\% &-5.49\%\\ \hline
Oct. 5, 2016& 14.73\% &8.57\% &-5.2\% &0.71\% &-9.71\% &-5.79\%\\ \hline
Oct. 6, 2016& 16.61\% &9.85\% &-3.07\% &1.22\% &-8.06\% &-5.21\%\\ \hline
Oct. 7, 2016& 15.08\% &8.47\% &-4.63\% &0.14\% &-9.81\% &-7.37\%\\ \hline
Oct. 8, 2016& 16.54\% &10.83\% &-4.62\% &-0.63\% &-7.35\% &-7.75\%\\ \hline
Oct. 9, 2016&  16.91\% &11.6\% &-4.49\% &0.39\% &-6.48\% &-7.02\%\\ \hline
Average     & 16.11\%	&9.94\%	&-4.42\%	&0.64\%	&-7.87\%	&-6.5\%\\ \hline
\multicolumn{7}{|c|}{Degradation of User Privacy With Respect to Proposed Hybrid Algorithm}\\ \hline
{Date}&$\omega =0$&$\omega =0.5$&$\omega =1$&MOIA&MOEA/D &NSGA-II\\ \hline
Oct. 3, 2016&  6.84\% &24.71\% &311.41\% &84.7\% &81.14\%	&36.99\%\\ \hline
Oct. 4, 2016& -5.43\% &7.43\%	 &235.56\% &53.61\% &67.51\%	 &20.78\%\\ \hline
Oct. 5, 2016&  8.11\% &32.29\% &276.448\% &91.4\% &106.99\% &49.12\%\\ \hline
Oct. 6, 2016& -5.797\% &14.92\% &260.49\% &75.13\% &98.02\% &35.11\%\\ \hline
Oct. 7, 2016& 5.08\% &17.38\% &254.18\%	&77.74\% &93.87\% &28.44\%\\ \hline
Oct. 8, 2016& 26.99\% &38.76\% &321.77\% &121.15\% &110.54\% &65.07\%\\ \hline
Oct. 9, 2016& -8.13\% &-3.77\% &216.24\% &68.07\% &57.96\% &37.79\%\\ \hline
Average     & 3.95\% &18.82\% &268.02\%	&81.69\% &88\% &39.04\%\\ \hline
\end{tabular}}
\end{table}

\begin{figure}
\hspace{-0.5cm}
  \includegraphics[width=10cm]{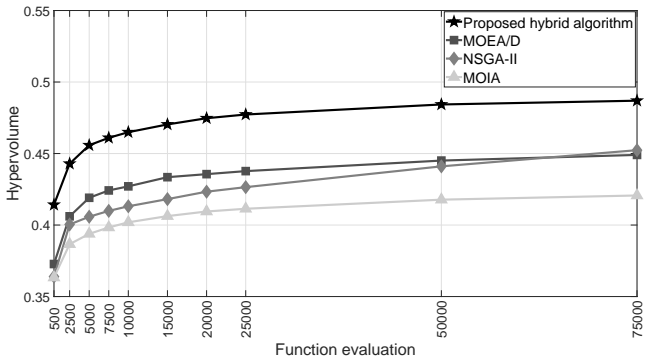}
  \caption{Convergence analysis of multiobjective evolutionary algorithms. }\label{fig_conv}
\end{figure}

\subsection{Analysis of Algorithm Convergence and Execution Time}

Fig.~\ref{fig_conv} presents the convergence analysis of multiobjective evolutionary algorithms.
The hypervolume serves as a performance index\cite{Conv_ana2014,HyperV_1,HyperV_2}. Because its absolute value depends on the chosen reference point,
algorithm convergence is assessed on the basis of its value variation and relative differences.
An algorithm is considered as convergent
when the corresponding hypervolume remains constant over function evaluation.
A higher hypervolume means a better level of convergence. In Fig.~\ref{fig_conv}, comparable algorithms converged when  approximately 25000 function evaluations were reached, implying that they had the same convergence rate; the proposed algorithm yielded a better level of convergence because a higher hypervolume was achieved.

It is known that single-objective optimization, i.e., weighted-sum methods in this study,  is faster than multiobjective optimization, i.e., the proposed hybrid algorithm, MOIA, MOEA/D, and NSGA-II.  As a reference for execution time, our experiments showed that proposed algorithm consumed 1.62
 seconds until the algorithm converged (25000 function evaluations) for power scheduling
while the weighted-sum methods consumed 1.57 seconds on average.
It is worth mentioning that  limited conclusions should be drawn here because
the execution time can be affected by several factors such as the used hardware, coding styles, and employed stopping criterion.
These factors make it difficult to provide a fair comparison.
Furthermore,  the execution time is not an important performance metric in our scenario because optimizing the power scheduling is done one day ahead on the basis of day-ahead pricing signals, which is an offline process.

\subsection{Analysis of Aggregate Behavior}

Algorithm performance was further evaluated in consideration of aggregate residential users.
For aggregate analysis, we considered a community of 500 residential users\cite{7586083} and adopted
the peak-to-average ratio (PAR)\cite{7575814}
\begin{equation}
PAR=\frac{ \underset{h\in \mathcal{H}}{\max} \;P^h }{\underset{h\in \mathcal{H}}\sum P^h /\left\vert \mathcal{H} \right\vert }
\label{eq_PAR}
\end{equation}
as the performance metric.
The lower the PAR, the better the result is.
For a smaller PAR, the associated power system is less affected by fluctuating power changes, leading to a more stable power system, and
the underlying power grid does not need to construct a large capacity to handle the peak power, drastically reducing implementation costs. The optimal PAR value is one, which means that the peak power consumption is equal to the average power consumption in the community.

\begin{figure}
  \centering
    \includegraphics[scale=0.18]{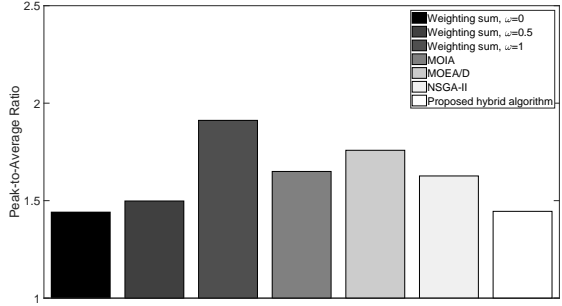}\\
  \caption{Peak-to-average ratio of 500 residential users in one week.}\label{fig_PAR}
\end{figure}

Fig.~\ref{fig_PAR} shows the resulting PAR of various methods. Excluding the weighted-sum method using  $\omega=0$, an undesirable method that only focused on user privacy while producing large energy costs, the proposed multiobjective approach yielded the best PAR. This indicates that our approach is scalable to  numbers of smart houses in a community.

In summary, the proposed hybrid algorithm provided a better balance between the energy cost and user privacy than existing approaches, with a comparable convergence rate and execution time.
The proposed approach was also scalable to power scheduling in a community of smart houses.
For single-objective optimization methods, the weighted-sum methods using $\omega=0$ and $\omega=1$  yielded large energy costs and significantly degraded user privacy, respectively; the weighted-sum method using $\omega=0.5$   found a balance between the two objectives, but was outperformed by our approach.
For existing multiobjective optimization methods,
algorithms  MOIA, MOEA/D, and NSGA-II were not robust in terms of privacy preservation.
Furthermore, the weighted-sum methods and MOEA/D used prior information for normalization, which may not be practical.

\section{Conclusion}\label{sec_conclusion}
In this study a power scheduling problem  of a smart house was investigated in which
the energy cost and privacy preservation were considered simultaneously.
A multiobjective  formulation was devised and a hybrid multiobjective algorithm was proposed.
Related numerical analysis was conducted to show several excellent properties of the proposed multiobjective methodology:
the proposed approach achieved a superior balance between the energy cost and user privacy
than existing energy management methods;
it had a comparable convergence rate while yielding a better level of convergence than the compared multiobjective evolutionary algorithms;
it resulted in a low peak-to-average ratio, which can be beneficial to the underlying power grid stability.

\appendix[Descriptions of Comparable Methods]
In general, candidate solutions for battery status $B(h)$ that satisfy the battery constraints are difficult to obtain
by random search. To address this difficulty, we calculated the upper and lower bounds on the next battery status by~(\ref{eq_Battery_bound}) and~(\ref{eq_Battery_next_status});
random trials within the associated ranges were then generated for twenty-four consecutive time slots. Finally, power storage profiles $S(h)$ were evaluated by~(\ref{eq_S_operate}).
This practice was adopted by  the following methods and algorithms.

For the weighted-sum method modified from\cite{7354489}, we solved
\begin{equation}
\begin{split}
&\mathop{\min}_{\substack{P_{b}^{h}, h_{c}, B(h)\\ h\in \mathcal{H},b\in \mathcal{A_{F}}\\c\in \mathcal{A_{S}}}}\omega \sum_{h\in \mathcal{H}}F_{Cost}+(1-\omega)F_{Privacy}\\\
&\text{subject to}~(\ref{eq_P_AF}),~(\ref{eq_Battery_bound}),~(\ref{eq_Battery_next_status})\text{ and}~(\ref{eq_S_and_PHA})
\end{split}
\end{equation}
using evolutionary algorithms for power scheduling in a smart home,
where $0 \leq \omega \leq 1$ denotes the weighting coefficient between  $F_{Cost}$ and $F_{Privacy}$.
It is worth mentioning that confining $\omega$ in the interval $[0,1]$ is mostly effective when the two objectives are normalized.
However, such normalization requires prior information.
In our simulations, the value of $\omega=0$ was set to solely address user privacy;
 $\omega=1$ was set to solely address the energy cost;
and $\omega=0.5$ with normalization in~(\ref{eq_normal_MOEAD}) was used to trade the privacy  off against the energy cost.
All these settings lead to a single-objective optimization problem.

For multiobjective optimization that addresses objectives in different dimensions, we employed algorithms MOIA, NSGA-II, and MOEA/D to solve corresponding MOPs.
The MOIA in\cite{7050260} was used to solve
\begin{equation}
\begin{split}
&\mathop{\min}_{\substack{P_{b}^{h}, h_{c}, B(h)\\ h\in \mathcal{H},b\in \mathcal{A_{F}}\\c\in \mathcal{A_{S}}}}F_{Cost}\qquad\qquad\\
&\mathop{\min}_{\substack{P_{b}^{h}, h_{c}, B(h)\\ h\in \mathcal{H},b\in \mathcal{A_{F}}\\c\in \mathcal{A_{S}}}}F_{Privacy}\qquad\qquad\\
&\text{subject to } (\ref{eq_P_AF}),~(\ref{eq_Battery_bound}),\text{ and}~(\ref{eq_Battery_next_status}).
\end{split}
\end{equation}
A constraint function
\begin{equation}
\begin{split}
U= \mathop{\max} \{S(h)-P_{HA}^{h},0\}
\end{split}
\end{equation}
derived from~(\ref{eq_S_and_PHA}) was used during the solving process to assess the feasibility of candidate solutions.
If a point is infeasible, then the associated $U$ would be greater than zero.
Only feasible points result in $U=0$.

To apply the NSGA-II\cite{6562791} and MOEA/D\cite{4358754}, a penalty method\cite{6336898} was employed to handle~(\ref{eq_S_and_PHA}). The power scheduling in a smart home was obtained by solving
\begin{equation}
\begin{split}
&\mathop{\min}_{\substack{P_{b}^{h}, h_{c}, B(h)\\ h\in \mathcal{H},b\in \mathcal{A_{F}}\\c\in \mathcal{A_{S}}}}F_{Cost}+k_1 U\\
&\mathop{\min}_{\substack{P_{b}^{h}, h_{c}, B(h)\\ h\in \mathcal{H},b\in \mathcal{A_{F}}\\c\in \mathcal{A_{S}}}}F_{Privacy}+k_1 U\\
&\text{subject to}~(\ref{eq_P_AF}),~(\ref{eq_Battery_bound})\text{ and}~(\ref{eq_Battery_next_status})
\end{split}
\end{equation}
where $k_1$ is a prescribed weighting coefficient. The value $k_1=10^{3}$ was set in our simulations.
For the MOEA/D, the objectives were normalized according to~(\ref{eq_normal_MOEAD}).

\end{document}